\documentclass[sigconf,screen]{acmart}
\AtBeginDocument{%
  \providecommand\BibTeX{{%
    \normalfont B\kern-0.5em{\scshape i\kern-0.25em b}\kern-0.8em\TeX}}}

\copyrightyear{2022} 
\acmYear{2022} 
\setcopyright{usgovmixed}\acmConference[WWW '22]{Proceedings of the ACM Web Conference 2022}{April 25--29, 2022}{Virtual Event, Lyon, France}
\acmBooktitle{Proceedings of the ACM Web Conference 2022 (WWW '22), April 25--29, 2022, Virtual Event, Lyon, France}
\acmPrice{15.00}
\acmDOI{10.1145/3485447.3511979}
\acmISBN{978-1-4503-9096-5/22/04}

\usepackage{epsfig}
\usepackage{tabularx}
\usepackage[ruled, linesnumbered]{algorithm2e}
\SetKwComment{Comment}{/* }{ */}
\usepackage{mathtools}
\usepackage{wrapfig}
\usepackage{colortbl}
\usepackage{multirow}
\usepackage{epstopdf} 
\usepackage{xspace}
\usepackage{bbm}

\usepackage{bbold}
\usepackage{graphicx}
\usepackage[labelformat=simple, subrefformat=parens, textfont=normal, labelfont=normal, list=true]{subcaption}

\usepackage{enumitem}
\usepackage{color}
\usepackage{graphicx,scalerel}

\newcommand{\otoprule}{\midrule[\heavyrulewidth]}

\newcommand{\eg}{{\it e.g.}}
\newcommand{\etal}{\textit{et al}.~}

\newcommand{\ie}{{\it i.e.}}

\newcommand{\sys}{\textit{Freyr}\xspace}

\newcommand{\paraspace}{\vspace{0.05in}}
\newcommand{\parab}[1]{\paraspace\noindent{\bf #1}}

\begin{document}

\title{Accelerating Serverless Computing by Harvesting Idle Resources}

\author{Hanfei Yu}
\email{hyu25@lsu.edu}
\affiliation{%
  \institution{Louisiana State University}
  \city{Baton Rouge}
  \state{LA}
  \country{USA}
}
\author{Hao Wang}
\email{haowang@lsu.edu}
\affiliation{%
  \institution{Louisiana State University}
  \city{Baton Rouge}
  \state{LA}
  \country{USA}
}
\author{Jian Li}
\email{lij@binghamton.edu}
\affiliation{%
  \institution{SUNY-Binghamton University}
  \city{Binghamton}
  \state{NY}
  \country{USA}
}
\author{Xu Yuan}
\email{xu.yuan@louisiana.edu}
\affiliation{%
  \institution{University of Louisiana at Lafayette}
  \city{Lafayette}
  \state{LA}
  \country{USA}
}
\author{Seung-Jong Park}
\email{sjpark@lsu.edu}
\affiliation{%
  \institution{Louisiana State University}
  \city{Baton Rouge}
  \state{LA}
  \country{USA}
}


\begin{abstract}


Serverless computing automates fine-grained resource scaling and simplifies the development and deployment of online services with stateless functions. 
However, it is still non-trivial for users to allocate appropriate resources due to various function types, dependencies, and input sizes. Misconfiguration of resource allocations leaves functions either under-provisioned or over-provisioned and leads to continuous low resource utilization.  
This paper presents \sys, a new resource manager (RM) for serverless platforms that maximizes resource efficiency by dynamically harvesting idle resources from over-provisioned functions to under-provisioned functions. \sys monitors each function's resource utilization in real-time, detects over-provisioning and under-provisioning, and learns to harvest idle resources safely and accelerates functions efficiently by applying deep reinforcement learning algorithms along with a safeguard mechanism. We have implemented and deployed a \sys prototype in a 13-node Apache OpenWhisk cluster. Experimental results show that 38.8\% of function invocations have idle resources harvested by \sys, and 39.2\% of invocations are accelerated by the harvested resources. \sys reduces the 99th-percentile function response latency by 32.1\% compared to the baseline RMs.


\end{abstract}

\ccsdesc[500]{Computer systems organization~Cloud computing}
\ccsdesc[300]{Computing methodologies~Planning and scheduling}

\keywords{Serverless computing, resource harvesting, reinforcement learning}

\maketitle

\section{Introduction}
%
The emergence of serverless computing has extensively simplified the way that developers access cloud resources. Existing serverless computing platforms, such as AWS Lambda, Google Cloud Functions, and Azure Functions, have enabled a wide spectrum of cloud applications, including web services~\cite{tanna2018serverless}, video processing~\cite{fouladi2017encoding, ao2018sprocket}, data analytics~\cite{jonas2017occupy, muller2020lambada}, and machine learning~\cite{carreira2019cirrus, siren2019infocom} with automated resource provisioning and management. 
By decoupling traditional monolithic cloud applications into inter-linked microservices executed by stateless \textit{functions}, serverless computing frees developers from infrastructure management and administration with fine-grained resource provisioning, auto-scaling, and pay-as-you-go billing~\cite{berkeley-view}.

Existing serverless computing platforms enforce \textit{static} resource provisioning for functions. For example, AWS Lambda allocates function CPU cores in a fixed proportion to the memory size configured by users~\cite{awslambdalimits}, leading to either CPU over-provisioned or under-provisioned for the function execution. 
Therefore, serverless service providers are enduring poor resource utilization due to users' inappropriate function configuration---some functions are assigned with more resources than they need~\cite{gunasekaran2020fifer}. The high concurrency and fine-grained resource isolation of serverless computing further amplify such inefficient resource provisioning.



A few recent studies attempted to address the above issues.  
Some researchers proposed to maximize resource utilization and reduce the number of cold-starts by predicting the keep-alive windows of individual serverless functions~\cite{fuerst2021faascache, azure-traces}. Fifer~\cite{gunasekaran2020fifer} incorporated the awareness of function dependencies into the design of a new resource manager to improve resource utilization. COSE~\cite{akhtar2020cose} attempts to use Bayesian Optimization to seek for the optimal configuration for functions. Furthermore, several works~\cite{cpu-cap, ensure, core-granular} aimed to accelerate functions and improve resource efficiency by adjusting CPU core allocations for serverless functions in reaction to their performance degradation during function executions. 


However, none of the existing studies has \textit{directly} tackled the low resource efficiency issue raised by the inappropriate function configurations. There are three critical challenges from the perspective of serverless service providers to address this issue.  \textit{First}, a user function is secured as a black box that shares no information about its internal code and workloads, making it hardly possible for the serverless system to estimate the precise resource demands of user functions. \textit{Second}, decoupling monolithic cloud applications to serverless computing architectures generates a variety of functions with diverse resource demands and dynamic input workloads. \textit{Third}, the resource provisioning for serverless functions is fine-grained spatially (\ie, small resource volumes) and temporally (\ie, short available time).


In this paper, we address the aforementioned challenges by presenting \sys, a new serverless resource manager (RM) that dynamically harvests idle resources to accelerate functions and maximize resource utilization. 
%
\sys estimates the CPU and memory saturation points respectively of each function and identifies whether a function is over-provisioned or under-provisioned.
For those over-provisioned functions, \sys harvests the wasted resources according to their saturation points; for those under-provisioned functions, \sys tries to accelerate them by offering additional, and just-in-need allocations to approach saturation points.
%
We apply an experience-driven algorithm to identify functions over-supplied and under-supplied by monitoring a series of performance metrics and resource footprints, including CPU utilization, memory utilization, and function response latency to estimate the actual resource demands of running functions. 
To deal with the highly volatile environment of serverless computing and large numbers of concurrent functions, we propose to apply the Proximal Policy Optimization (PPO) algorithm~\cite{ppo2} to learn from the realistic serverless system and make per-invocation resource adjustments.
Besides, we design a safeguard mechanism for safely harvesting idle resources without introducing any performance degradation to function executions that have resource harvested.  
%
%

%
We implement \sys based on Apache OpenWhisk~\cite{openwhisk}, a popular open-source serverless computing platform. We develop a Deep Reinforcement Learning (DRL) model and training algorithm using PyTorch and enable multi-process support for concurrent function invocations. 
We evaluate \sys with the other three baselines on an OpenWhisk cluster using realistic serverless workloads. 
Compared to the default resource manager in OpenWhisk, \sys reduces the 99th-percentile function response latency of invocations\footnote{In this paper, a function denotes an executable code package deployed on serverless platforms, and a function invocation is a running instance of the code package.} by 32.1\%. Particularly, \sys harvests idle resources from 38.8\% of function invocations while accelerating 39.2\% on the OpenWhisk cluster. Notably, \sys only degrades a negligible percentage of function performance under the system performance variations of the OpenWhisk cluster.


\section{Background and Motivation}
This section first introduces the status quo of resource provisioning and allocation in serverless computing. Then, we use real-world experiments to demonstrate that serverless functions can easily become under-provisioned or over-provisioned, and motivate the necessity to accelerate under-provisioned functions and optimize resource utilization by harvesting idle resources at runtime. 




\subsection{Resource Provisioning and Allocation in Serverless Computing}
Existing serverless computing platforms (\eg, AWS Lambda, Google Cloud Functions, and Apache OpenWhisk) request users to define memory up limits for their functions and allocate CPU cores according to a fixed proportion of the memory limits~\cite{peeking-behind-serverless,openwhisk,awslambda,gcfunction}. Obviously, the fixed proportion between CPU and memory allocations leaves serverless functions either under-provisioned or over-provisioned because functions' CPU and memory demands differ significantly.
%

Further it is non-trivial for users to accurately allocate appropriate amounts of resource for their functions~\cite{ensure, akhtar2020cose} due to various function types, dependencies, and input sizes.  Users are prone to oversize their resource allocation to accommodate potential peak workloads and failures~\cite{ensure, iorgulescu2018perfiso}. 
Finally, users' inappropriate resource allocations and providers' fixed CPU and memory provisioning proportion jointly degrade the resource utilization in serverless computing as resources allocated to functions remain idle (more discussion in Supplementary Materials~\ref{sup:deploy-sys}).   

\subsection{Resource Saturation Points}
\label{subsec:saturation}

We further demonstrate how easily a serverless function becomes under-provisioned or over-provisioned by introducing a new notion of \textit{saturation points}.
Given a function and an input size, there exists a resource allocation \textit{saturation point}---allocating resource beyond this point can no longer improve the function's performance, but allocating resource below this point severely degrades the performance.

We profile the saturation points of two applications: email generation (EG) and K-nearest neighbors (KNN), representing two popular serverless application categories: web applications and machine learning, respectively.
We identify the allocation saturation points of CPUs and memory separately by measuring the response latency of functions allocated with different number of CPU cores and different sizes of memory. When adjusting a function's CPU (memory) allocation, we fix its memory (CPU) allocation to 1,024 MB (8 cores). 

\begin{figure}[t]
    \centering
    \includegraphics[width=.9\linewidth]{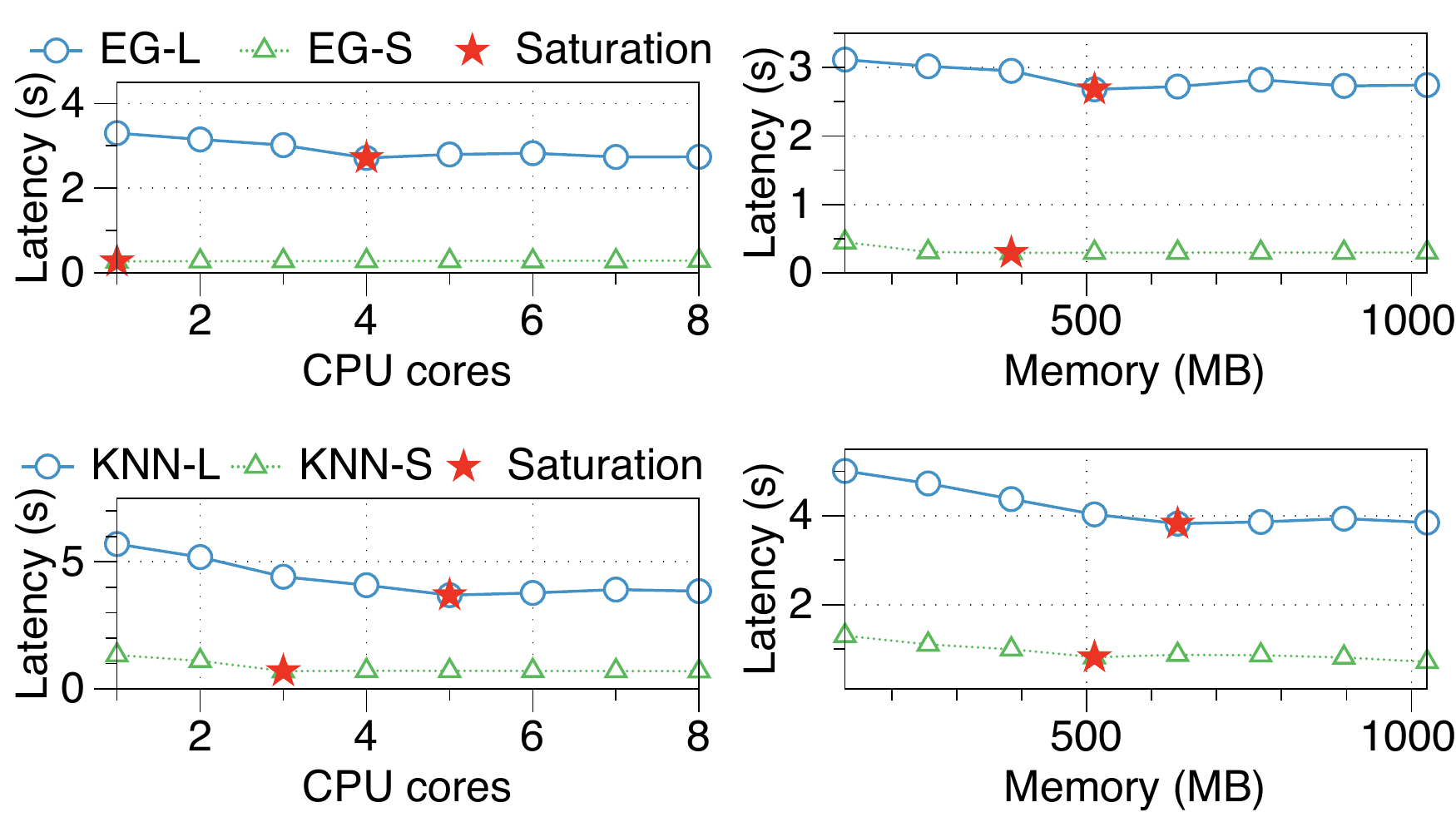}
    \Description{}
    \vspace{-0.15in}
    \caption{Saturation points of EG and KNN with small (S) and large (L) workload sizes. EG-S (L) generates 1K (10K) emails, and KNN-S (L) inputs 2K (20K) data samples.}
    \label{fig:motivation-saturation}
    \vspace{-0.2in}
\end{figure}

Figure~\ref{fig:motivation-saturation} shows that saturation points vary from functions and input sizes. It is non-trivial for users to identify the saturation points for every function with specific input sizes in their applications. Particularly, serverless functions are typically driven by events with varying input sizes. Without dynamic and careful resource allocations, functions tend to become either over-provisioned or under-provisioned. 
%

\subsection{The Need for Harvesting Idle Resources}
\label{sec:motivating}

%


Resource harvesting is a common methodology in virtual environments that increases resource utilization by reallocating idle resources to under-provisioned services without degrading the performance of services being harvested~\cite{ambati2020providing, yanqizhang2021sosp, wang2021smartharvest}. 

To motivate the need for dynamic resource harvesting in serverless computing, we compare the function response latency achieved by the default resource manager (Fixed RM) and greedy resource manager (Greedy RM) when executing four real-world serverless functions. 
The Fixed RM simply accepts and applies a fixed resource allocation pre-defined by users, such as the RM in OpenWhisk and AWS Lambda. 
The Greedy RM dynamically harvests CPU cores from functions over-provisioned and assigns the harvested CPU cores to functions under-provisioned in a first-come-first-serve manner based on the estimated function saturation points learned from functions' recent resource utilization (details in Section~\ref{sec:evaluation}). In this experiment, we collect historical resource utilizations of four functions and profile their saturation points.

\begin{figure}[t]
    \captionsetup[subfigure]{aboveskip=-1pt,belowskip=-1pt}
    \centering
    \begin{subfigure}[b]{0.22\textwidth}
        \centering
        \includegraphics[width=\textwidth]{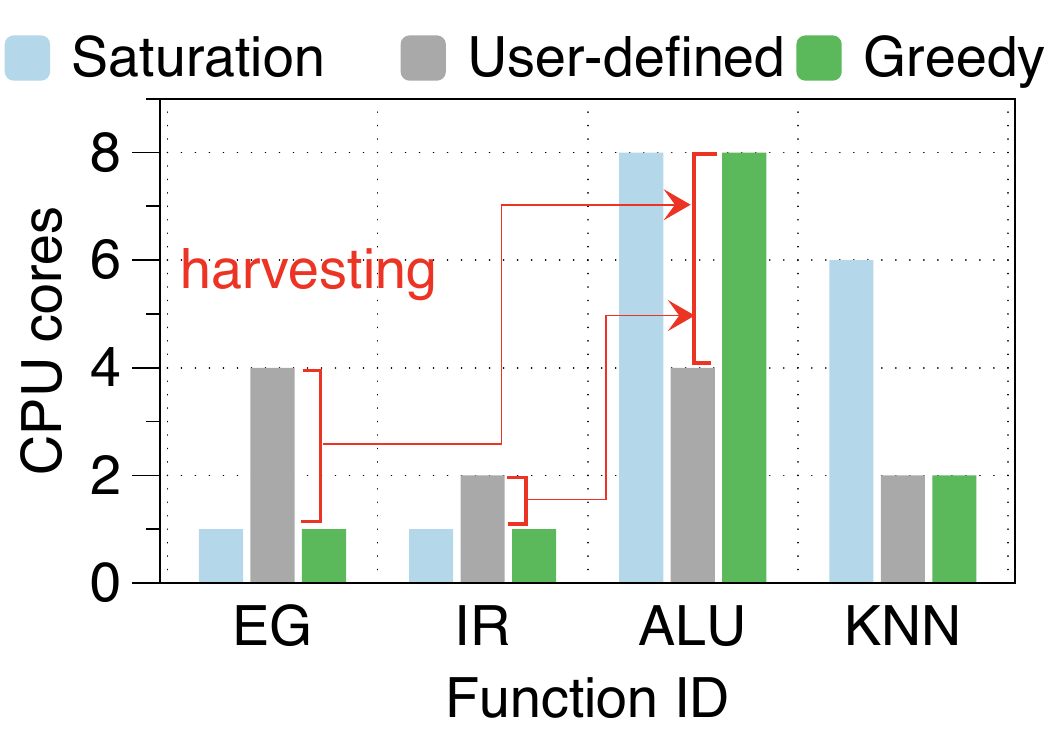}
        \caption{CPU allocation}
        \Description{}
        \label{subfig:motivation_alloc}
    \end{subfigure}
    \begin{subfigure}[b]{0.22\textwidth}
        \centering
        \includegraphics[width=\textwidth]{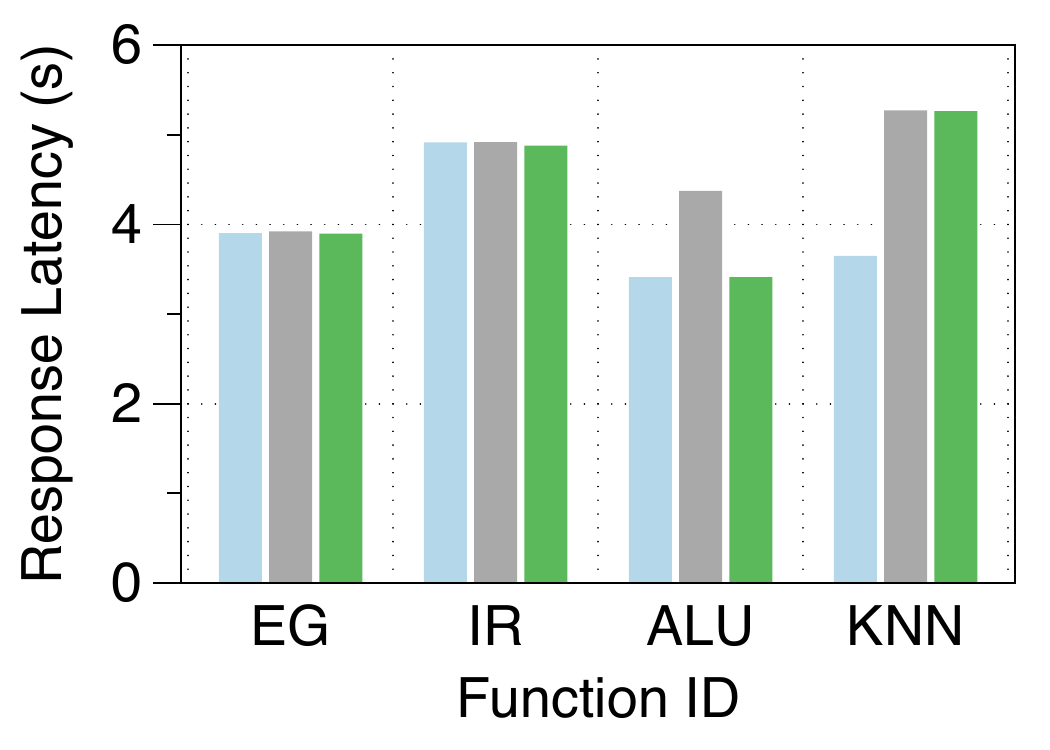}
        \caption{Function response latency}
        \Description{}
        \label{subfig:motivation_latency}
    \end{subfigure}
    \vspace{-0.1in}
    \caption{The CPU allocation and response latency of four real-world functions: EG, image recognition (IR), arithmetic logic units (ALU), and KNN, where the EG generates 100K emails, the IR classifies ten images, the ALU calculates 20M loops, and the KNN inputs 20K data samples.}
    \label{fig:motivation}
    \vspace{-0.2in}
\end{figure}

Figure~\ref{subfig:motivation_alloc} shows the Greedy RM accelerates the ALU by harvesting three CPU cores from the EG (\ie, the EG function invocation) and one CPU core from the IR. Though the KNN is also under-provisioned, the Greedy RM assigns all harvested CPU cores to the ALU since the ALU is invoked before the KNN. As a comparison, Figure~\ref{fig:motivation} also plots the saturation points of each function invocation and their response latency when allocated with saturated resources.
Figure~\ref{subfig:motivation_latency} shows the Greedy RM can increase resource utilization and accelerate under-provisioned functions without sacrificing over-provisioned functions' performance in the motivation scenario.

\subsection{Deep Reinforcement Learning} 
\label{subsec:drl}

Due to the volatility and burstiness of serverless computing, it is non-trivial to accurately estimate the saturation points based on functions' recent resource utilization, and the greedy resource harvesting and re-assignment can hardly minimize the overall function response latency. Thus, we propose to utilize reinforcement learning (RL) algorithms to learn the optimal resource harvesting and re-assignment strategies. 

At every timestep $t$, the agent is in a specific state $s_t$, and evolves to state $s_{t+1}$ according to a Markov process with the transition probability $\mathbb{P}(s_t,a_t,s_{t+1})$ when action $a_t$ is taken~\cite{sutton1998reinforcement}.  The immediate reward for the agent to take action $a_t$ in state $s_t$ is denoted as $r_t.$ The goal of the agent is to find a policy $\pi$ that makes decisions regarding what action to take at each timestep, $a_t \sim \pi(\cdot|s_t)$, so as to maximize the expected cumulative rewards, $\mathbb{E}_\pi[\sum^{\infty}_{t=1} \gamma^{t-1}r_t]$, where $\gamma \in (0,1]$ is a discount factor. 

To capture the patterns of real-world systems and address the curse-of-dimensionality, deep reinforcement learning (DRL) has been introduced to solve scheduling and resource provisioning problems in distributed systems~\cite{deeprm, decima, mondal2021scheduling, li2021george}, where deep neural networks serve as the \textit{function approximators} that describe the relationship between decisions, observations, and rewards.

\section{Overview}
\label{sec:overview}



\subsection{Design Challenges}
\label{subsec:challenges}


Unlike long-running VMs with substantial historical traces for demand prediction and flexible time windows for resource harvesting, function executions in serverless computing are highly concurrent, event-driven, and short-lived with bursty input workloads~\cite{datadogserverless}, making it hardly practical to reuse the existing VM resource harvesting methods. 
To enable efficient and safe resource harvesting and performance acceleration in serverless computing, \sys's design tackles three key challenges:

\textbf{Volatile and bursty serverless environments.} The heterogeneity of serverless functions, the high concurrency of invocation events, and the burstiness of input workloads jointly make it non-trivial to accurately determine whether a function execution has idle resources to be harvested. Besides, serverless functions are sensitive to the latency introduced by resource harvesting and re-assignment due to their short lifetime and event-driven nature. 


\textbf{Huge space of harvesting and re-assignment decisions.} 
Unlike the default resource managers that enforce a fixed proportion between the CPU and memory allocations, 
we decouple the resource provisioning for CPU and memory for more accurate resource harvesting and re-assignment, leading to a two-dimensional resource pool for \sys to seek for the optimal resource allocation. This is an immense action space for the DRL agent.  
For example, AWS Lambda allows any memory sizes between 128 MB and 10,240 MB and up to 6 CPU cores---60,672 choices in total. Such a huge action space complicates the DRL algorithm design and extensively increases the computation complexity to train the DRL agent. 

\textbf{Potential performance degradation.} While \sys harvests resources from functions deemed as over-provisioned and improves the entire workload, one necessary requirement is to prevent the performance of those functions from degrading. It is vital to guarantee Service Level Objectives (SLOs) of each individual function, \ie, harvested functions have no significant performance degradation.


\subsection{\sys's Architecture}

\begin{figure}[t]
  \centering
  \includegraphics[width=0.3\textwidth]{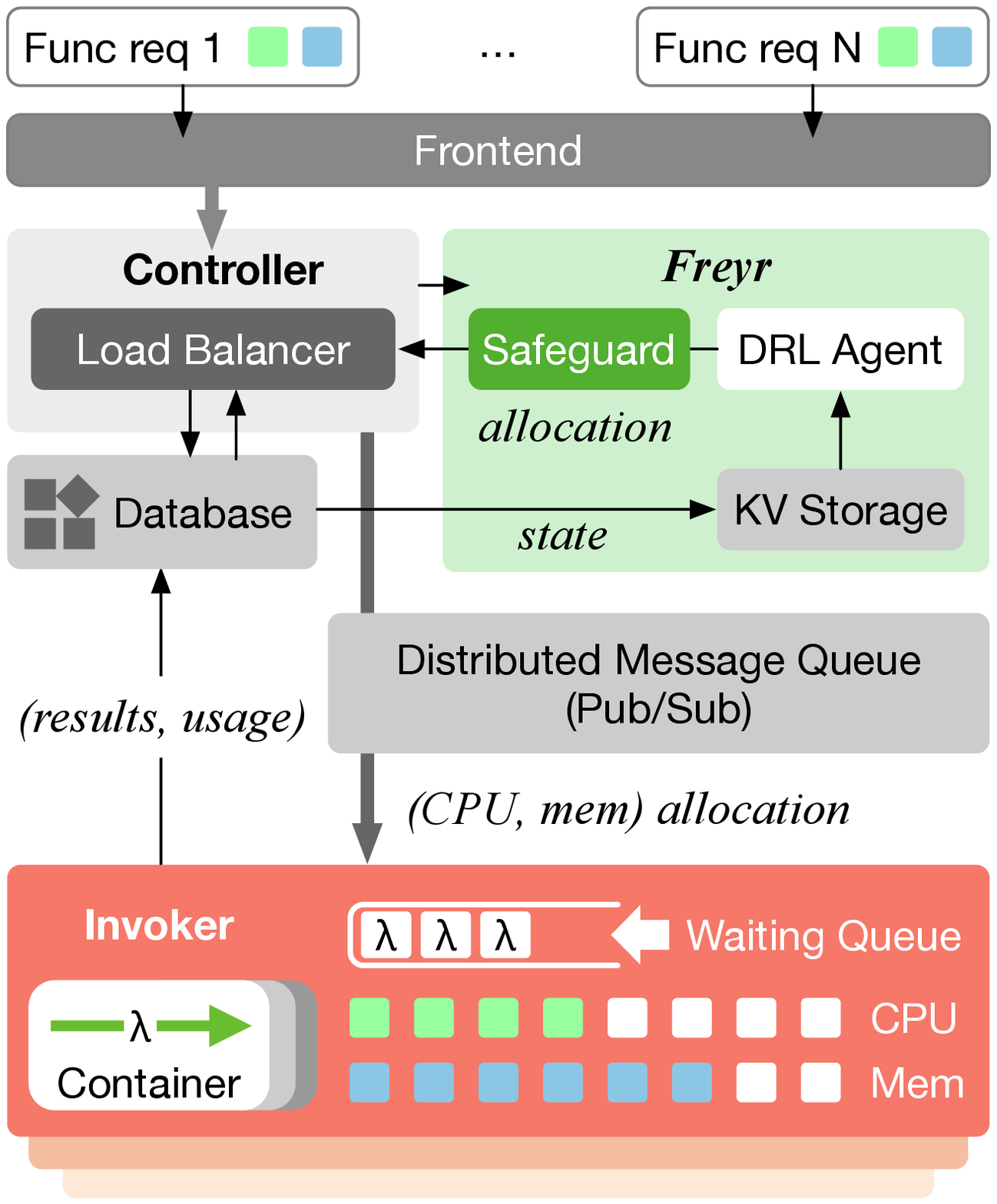}
  \vspace{-0.15in}
  \caption{ \sys's architecture.}
  \Description{}
  \vspace{-0.2in}
  \label{fig:openwhisk}
\end{figure}

\sys is a resource manager in serverless platforms that dynamically harvests idle resources from over-provisioned function invocations and reassign the harvested resources to accelerate under-provisioned function invocations. 
It is located with the controller of a serverless computing framework and interacts with the container system (\eg, Docker~\cite{docker}) that executes function invocations. 

Figure~\ref{fig:openwhisk} shows an overview of \sys's architecture. 
First, concurrent function requests arrive at the frontend to invoke specific functions with user-defined resource allocations. 
The controller admits the function requests, registers their configurations, and schedules them to the invokers. 
Before the execution of functions, \sys inputs observations from serverless platform database and makes resource harvesting and re-assignment decisions. 
The controller instructs invokers to enforce the decisions when executing function invocations. 

To handle the \textbf{volatile and bursty serverless environments}, \sys is designed to be event-driven with multi-process support that the arrival of a function request triggers \sys to make resource harvesting decisions. 
To shrink the \textbf{huge decision space}, \sys trains a score network to justify each allocation option of function invocations, converting the action space of all potential allocation options to a score for individual allocation option. 
\sys evaluates the score of each allocation option using this score network and enforces the allocation option with the highest score. 
%
To avoid \textbf{potential performance degradation} of functions with resources harvested, \sys applies a \textit{safeguard mechanism} to prevent those potentially dangerous allocation options and guarantees the SLOs of every function invocation within a workload.
The safeguard examines whether the allocation decision made by the DRL agent is below a function's historical resource usage peak. 
Besides, the safeguard monitors the function's runtime resource utilization and returns all harvested resources by calling a \textit{safeguard invocation} when its resources appear to be fully utilized.

\section{Design}


\subsection{Problem Formulation}
\label{subsec:formulation}



We consider a serverless platform that handles a workload $W$ with multiple concurrent function invocations. 
Let $f$ denote a function invocation in $W$.
We assume the response latency $e$ of $f$ is dominated by CPU and memory.
Each function invocation $f$ has a resource allocation $p = (p_c, p_m)$, where $p_c$ and $p_m$ denote a set of CPU and memory resources, respectively.
We assume $p$ is non-preemptive and fixed when the platform is executing $f$, \ie, $p$ is consistently provisioned to $f$ until the execution completes.
Thus, we define the relationship between the response latency and the resource allocation as: $e = B(p)$.
Section~\ref{subsec:saturation} demonstrates that a function invocation has a pair of saturation points for CPU and memory denoted by $p^\Xi = (p^\Xi_c, p^\Xi_m)$, respectively.

The platform determines whether it can harvest or accelerate a function invocation $f$ by comparing $p$ with $p^\Xi$: if $p^\Xi_c < p_c$ ($p^\Xi_m < p_m$), $f$ has idle CPU (memory), the platform can harvest at most $p_c - p^\Xi_c$ resources without increasing response latency $e$; if $p^\Xi_c > p_c$ ($p^\Xi_m > p_m$), the allocation of $f$ hasn't saturated, the platform can provide $f$ with at most $p^\Xi_c - p_c$ resources to improve the performance of $f$, \ie, reduce response latency $e$.
Thus for CPU or memory, function invocations in a workload $W$ can be classified into three groups of invocations: $W = W_h + W_a + W_d$, where $W_h$ denotes the set of invocations that can be harvested, $W_a$ denotes the set of invocations that can be accelerated, and $W_d$ denotes the set of invocations which have descent user configurations ($p^\Xi = p$).

\begin{figure*}[t]
    \centering
    \includegraphics[width=0.8\textwidth]{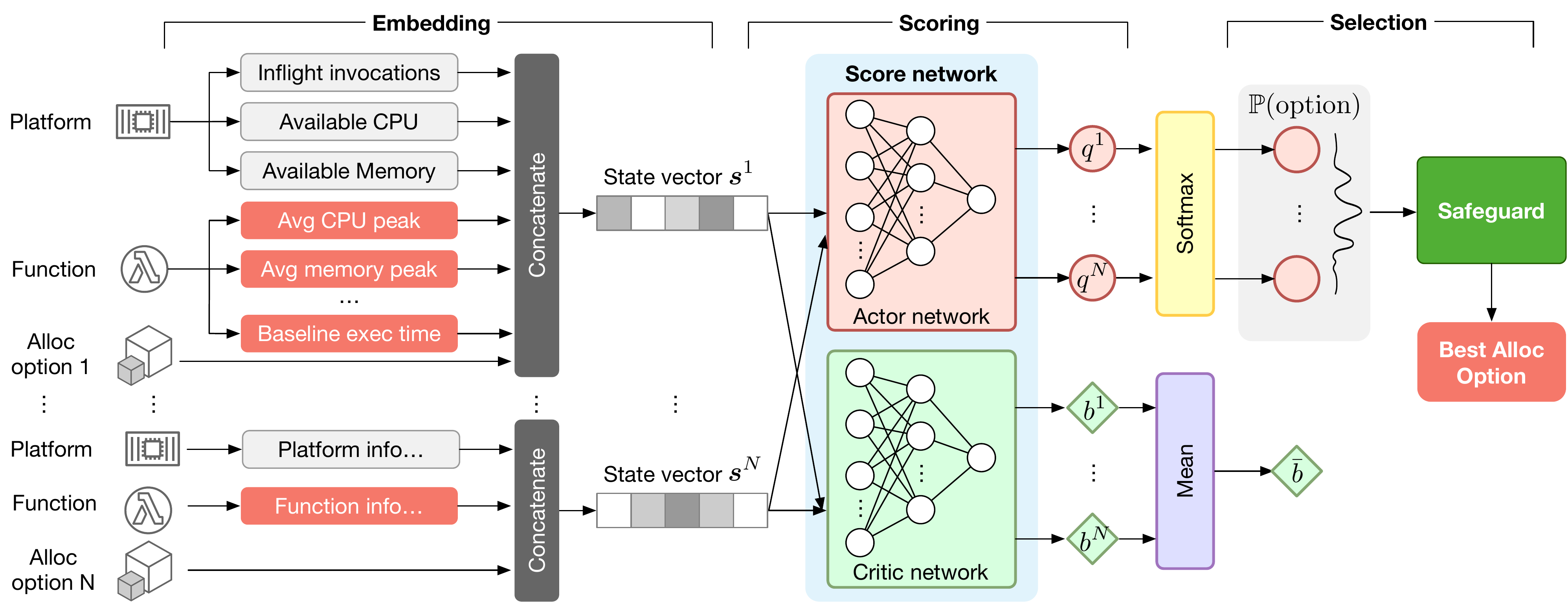}
    \vspace{-0.1in}
    \caption{The workflow of \sys.}
    \vspace{-0.1in}
    \label{fig:workflow}
\end{figure*}

We define a \textit{slowdown} value as the performance metric to avoid prioritizing long invocations while keeping short invocations starving.
Recall that $W$ denotes the workload, $f$ denotes a function invocation in $W$. 
Function invocations arrive at the platform in a sequential order.
At the first invocation of a function, the platform captures the response latency $e^b$ with resources $(p^b_c, p^b_m)$ configured by the user and employs it as a baseline denoted by $b$. 
When $i$-th invocation completes execution, the platform captures the response latency $e^i$ of it. 
The slowdown of the $i$-th invocation is calculated as
\begin{equation}
    \textit{slowdown} := \frac{e^i}{e^b}.
    \label{eq:rfet}
\end{equation}
We normalize the response latency of each invocation with baseline latency of user configuration.
Intuitively, the slowdown indicates how a function invocation performs regardless of its duration length.
A function invocation may be accelerated while being harvested at the same time (\eg, $p^\Xi_c < p_c$ while $p^\Xi_m > p_m$).
In this case, the slowdown is a mixed result.
For individual invocations, we only focus on the performance regardless of details of resource allocation, \ie, the invocation is good as long as it yields low slowdown.
We use average slowdown to measure how well a workload is handled by the platform with harvesting and acceleration.
Hence, the goal is to find a set of resource allocation $p = (p^1, p^2, ..., p^{|W|})$ which minimizes the average slowdown of a workload, defined as
\begin{equation}
    avg\_slowdown := \frac{1}{|W|}\sum_{f \in W} \frac{e^i}{e^b}
            = \frac{1}{|W|}\sum_{f \in W} \frac{B(p^i)}{B(p^b)}
\end{equation}
However, as introduced in Section~\ref{subsec:saturation}, estimating varying saturation points of sequential function invocations posts a challenging sequential decision problem.
The complex mapping from set of $p$ to objective average slowdown can hardly be solved by existing deterministic algorithms.
Hence, we opt for DRL and propose \sys, which learns to optimize the problem by replaying experiences through training.
\sys observes information from platform level and function level in real time.
Figure \ref{fig:workflow} depicts how \sys estimates CPU/memory saturation points. Given a function invocation, we encode every possible CPU and memory option into a scalar value representing the choice.


\subsection{Information Collection and Embedding}
\label{subsec:embedding}

When allocating resources for a function invocation, \sys collects information from two levels: platform level and function level, as summarized in Table~\ref{table:state-space}. Specifically, for the platform, \sys captures the number of invocations remaining in the system (\ie, \texttt{inflight\_request\_num}), available CPU cores, and available memory. For the incoming function, \sys queries invocation history of the function which records average CPU peak, average memory peak, average inter-arrival time (IAT), average execution time, and baseline execution time (\ie, \texttt{baseline}) with user-requested resources.

Once collecting such information, \sys encapsulates them with a potential resource allocation option.  More precisely, we embed information and the potential configuration option together into a flat state vector as input to \sys agent, with the information embedding process illustrated in Figure~\ref{fig:workflow}.

\subsection{Score Network}
\label{subsec:score-network}

\sys uses a \textit{score network} to calculate the priority of selecting potential resource allocation options. Figure~\ref{fig:workflow} visualizes the policy network of \sys agent, and illustrates the workflow of how the agent selects the best allocation option based on states.
At time $t$, a function invocation arrives at the platform which has in total $N$ potential resource configuration options. After embedding procedure, \sys collects a batch of state vectors $\boldsymbol{s}_t = (\boldsymbol{s}^1_t, \ldots, \boldsymbol{s}^n_t, \ldots, \boldsymbol{s}^N_t)$, where $\boldsymbol{s}^n_t$ maps the state to the $n$-th option. \sys inputs $s_t$ to the score network. We implement the score network using two neural networks, an \textit{actor network} and a \textit{critic network}. The actor network computes a score $q^n_t$, which is a scalar value mapped from the state vector $\boldsymbol{s}^n_t$ representing the priority to select configuration option $n$. Then \sys applies a Softmax operation to the scores $(q^1_t, \ldots, q^n_t, \ldots, q^N_t)$ to compute the probability of selecting option $n$ based on the priority scores, given by
\begin{equation*}
    \mathbb{P}_t(\textit{option}=n) = \frac{\exp(q^n_t)}{\sum^N_{n=1} \exp(q^n_t)},
    \label{eq:softmax}
\end{equation*}
at time $t$.
The critic network outputs a baseline value $b^n_t$ for option $n$, the average baseline value $\Bar{b}_t$ is calculated as
\begin{equation}
    \Bar{b}_t = \frac{1}{N}\sum^N_{n=1} b^n_t,
    \label{eq:baseline-value}
\end{equation}
which is used to reduce variance when training \sys. The whole operation of policy network is end-to-end differentiable.

The score network itself contains no manual feature engineering. \sys agent automatically learns to compute accurate priority score of allocation options through training. More importantly, \sys uses the same score network for all function invocations and all potential resource allocation options. By embedding options into state vectors, \sys can distinguish between different options and use the score network to select the best option. Reusing the score network reduces the size of networks and limits the action space of \sys agent significantly.

\begin{table}[tb]
    \centering
    \caption{The observation state space of the DRL agent.}
    \vspace{-0.1in}
    \begin{small}
        \begin{tabular}{ll}
            \toprule%
            \multirow{2}{4em}{\textbf{Platform State}} & \texttt{avail\_cpu}, \texttt{avail\_mem}               \\
                                                       & \texttt{inflight\_request\_num}                        \\\midrule%
            \multirow{3}{4em}{\textbf{Function State}} & \texttt{avg\_cpu\_peak}, \texttt{avg\_mem\_peak},      \\
                                                       & \texttt{avg\_interval}, \texttt{avg\_execution\_time}, \\
                                                       &\texttt{baseline}          \\
            \bottomrule%
        \end{tabular}
    \end{small}
    \vspace{-0.25in}
    \label{table:state-space}
\end{table}

\subsection{Safeguard}
\label{subsec:safeguard}

We design \sys to improve both over-provisioned and under-provisioned functions. However, when harvesting resources from functions deemed as over-provisioned, it is possible that \sys under-predicts their resource demands. The performance of functions degrades when being over-harvested. We devise a safeguard mechanism atop \sys to regulate decisions by avoiding decisions that may harm performance and returning harvested resources immediately when detecting a usage spike. We use this safeguard mechanism to mitigate obvious performance degradation of individual functions.

\begin{algorithm}[thb]
\DontPrintSemicolon
\SetNoFillComment
    \caption{Safeguard mechanism atop \sys.}
    \label{algo:safeguard}
    \While{request\_queue.notEmpty}{
        function\_id $\gets$ request\_queue.dequeue()\;
        calibrate\_baseline $\gets$ False\;
        last\_request $\gets$ \texttt{QueryRequestHistory}(function\_id)\;
        \eIf{last\_request == None}{
            \tcc*[h]{Trigger safeguard}\;
            range $\gets$ [user\_defined]\;
            calibrate\_baseline $\gets$ True\;
        }{
            threshold $\gets$ 0.8\;
            last\_alloc $\gets$ last\_request.alloc\;
            last\_peak $\gets$ last\_request.peak\;
            recent\_peak $\gets$ \texttt{GetRecentPeak}(function\_id)\;
            \eIf{last\_peak $<$ user\_defined}{ 
                \tcc*[h]{Over-provisioned}\;
                \eIf{last\_peak $/$ last\_alloc $\geq$ threshold}{
                    \tcc*[h]{Trigger safeguard}\;
                    range $\gets$ [user\_defined]\;
                    calibrate\_baseline $\gets$ True\;
                }{
                    range $\gets$ [recent\_peak + 1, user\_defined]\;
                }
            }{  \tcc*[h]{Under-provisioned}\;
                
                range $\gets$ [recent\_peak + 1, max\_per\_function]\;
            }
        }
        alloc\_option $\gets$ \texttt{\sys}(function\_id, range)\;
        \texttt{Invoke}(function\_id, alloc\_option, calibrate\_baseline)\;
    }
\end{algorithm}

Algorithm~\ref{algo:safeguard} summarizes the safeguard mechanism built atop \sys. 
We refer safeguard invocation as invoking the function with user-defined resources. 
When there are no previous invocations, \sys triggers the safeguard to obtain resource usage and calibrate the baseline mentioned in Equation~\ref{eq:rfet} (lines 5--7). 
For further invocations, \sys queries the history of function and polls the usage peak, allocation of the last invocation, and the highest peak since last baseline calibration (lines 10--12). 
\sys first checks current status of the function, \ie, over-provisioned or under-provisioned (line 13). 
We assume functions with resource usage below 80\% of user-requested level is over-provisioned. 
For over-provisioned (harvested) functions, \sys then checks the usage peak of last invocation (line 14). 
If the usage peak approaches 80\% of allocation, we suspect there may be a load spike, which could use more resources than current allocation. 
This triggers the safeguard invocation and baseline re-calibration, \sys immediately returns harvested resource to the function at the next invocation (lines 15--16). 
If there is no usage spike, \sys is allowed to select an allocation option from recent peak plus one unit to a user-requested level (line 18). 
For under-provisioned functions, \sys is allowed to select from recent peak plus one unit to the maximum available level (line 21). 
After an allocation option is selected, \sys invokes the function and forwards the invocation to invoker servers for execution.

Supplementary Materials \ref{sup:deep_dive} presents a sensitivity analysis of safeguard thresholds and shows that the safeguard mechanism effectively regulates decisions made by \sys and protects SLOs of functions that have resources harvested.

\subsection{Training the DRL Agent}
\label{subsec:training}

\sys training proceeds in \textit{episodes}.
In each episode, a series of function invocations arrive at the serverless platform, and each requires a two-dimensional action to configure CPU and memory resources.
When the platform completes all function invocations, the current episode ends.
Let $T$ denote the total number of invocations in an episode, and $t_i$ denote the arrival time of the $i$-th invocation.
We continuously feed \sys with a reward $r$ after it takes an action to handle an invocation.
Concretely, we penalize \sys with
\begin{equation*}
    r_i = - \sum_{f \in S|^{t_i}_{t_{i-1}}} \frac{e^i}{e^b} + R_{(\textit{slowdown}<1)} - R_{(\textit{slowdown}>1)},
\end{equation*}
after taking action on the $i$-th invocation, where $W$ is the set of invocations that finish during the interval $[t_{i-1}, t_i)$, $\frac{e^i}{e^b}$ is the slowdown of an invocation $f$ introduced in Section~\ref{subsec:formulation}, and two constant summaries for awarding good and penalizing bad actions ($R_{(\textit{slowdown}<1)}$ and $R_{(\textit{slowdown}>1)}$).  
The goal of the algorithm is to maximize the expected cumulative rewards given by
\begin{equation}
    \mathbb{E}\Bigg[\sum^T_{i=1}\gamma^{t-1} \Bigg( - \sum_{f \in S|^{t_i}_{t_{i-1}}} \frac{e^i}{e^b} + R_{(\textit{slowdown}<1)} - R_{(\textit{slowdown}>1)}\Bigg)\Bigg].
    \label{eq:rewards}
\end{equation}
Similar to \cite{deeprm}, we set the discount factor $\gamma$ in Equation \ref{eq:rewards} to be 1.
Hence, \sys learns to minimize the overall \textit{slowdown} of the given workload. 

\begin{figure}[bt]
    \captionsetup[subfigure]{aboveskip=-1pt,belowskip=-1pt}
    \centering
    \begin{subfigure}[b]{0.21\textwidth}
        \centering
        \includegraphics[width=\textwidth]{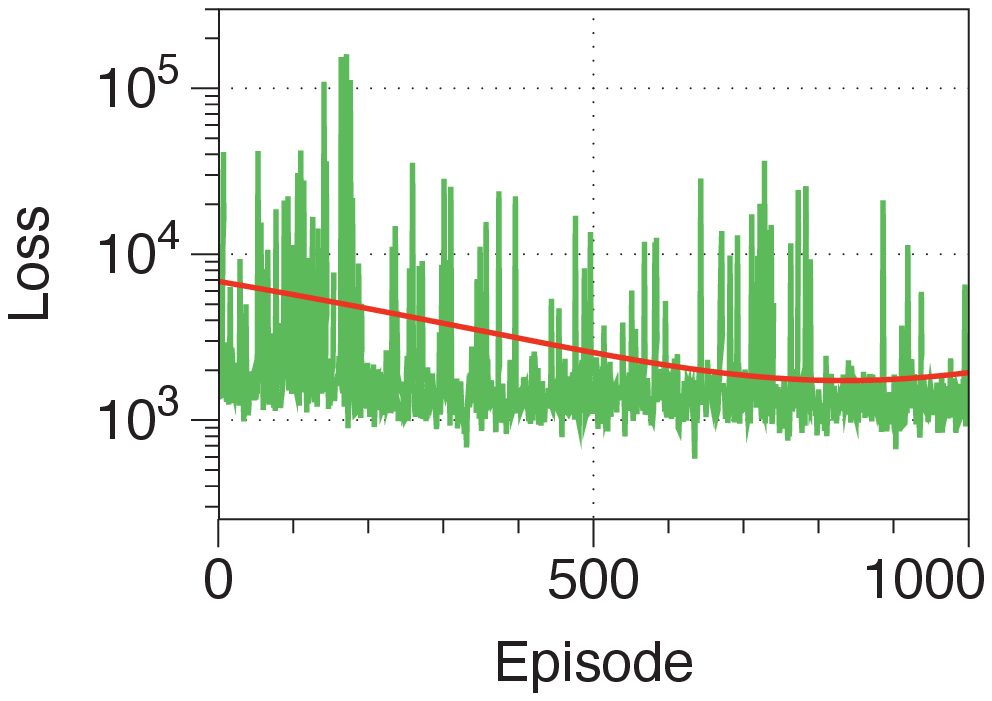}
        \caption{Cumulative average loss}
        \label{subfig:openwhisk_loss}
    \end{subfigure}
    \begin{subfigure}[b]{0.21\textwidth}
        \centering
        \includegraphics[width=\textwidth]{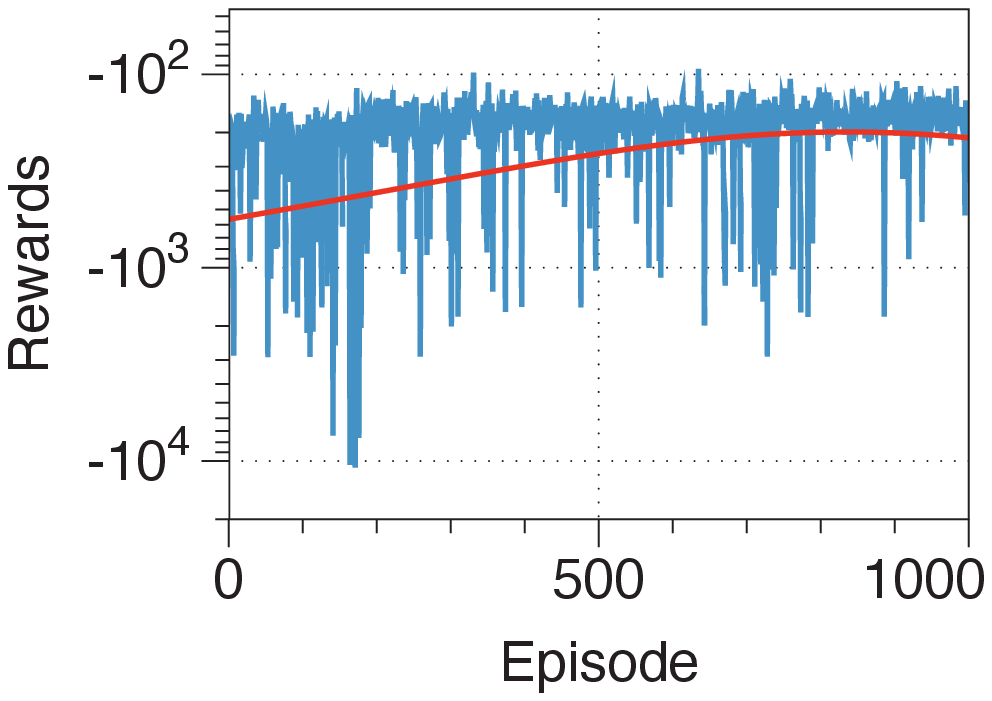}
        \caption{Cumulative rewards}
        \label{subfig:openwhisk_rewards}
    \end{subfigure}
    \vspace{-0.1in}
    \caption{The trends of cumulative average loss (left) and cumulative rewards (right) of \sys's 1,000-episode training on the OpenWhisk testbed.}
    \vspace{-0.15in}
    \label{fig:openwhisk_training}
\end{figure}

We use the algorithm~\ref{algo:freyr} to train \sys with 4 epochs per surrogate optimization and a 0.2 clip threshold \cite{ppo2}. 
We update the policy network parameters using the AdamW optimizer \cite{adamw} with a learning rate of 0.001. 
We train \sys with 1,000 episodes. 
The total training time is about 120 hours. 
Figure~\ref{fig:openwhisk_training} shows the learning curve and cumulative rewards of \sys training on OpenWhisk testbed. 
In Figure~\ref{subfig:openwhisk_loss}, the descending loss trendline indicates that \sys gradually learns to make good resource management decisions.
In Figure~\ref{subfig:openwhisk_rewards}, the ascending trendline shows that \sys seeks to maximize the cumulative rewards through training. Supplementary Material~\ref{sup:training} and \ref{sup:openwhisk} introduces the details of \sys's training algorithm and implementation.

\section{Evaluation}
\label{sec:evaluation}

\begin{figure*}[t]
    \centering
    \includegraphics[width=0.76\textwidth]{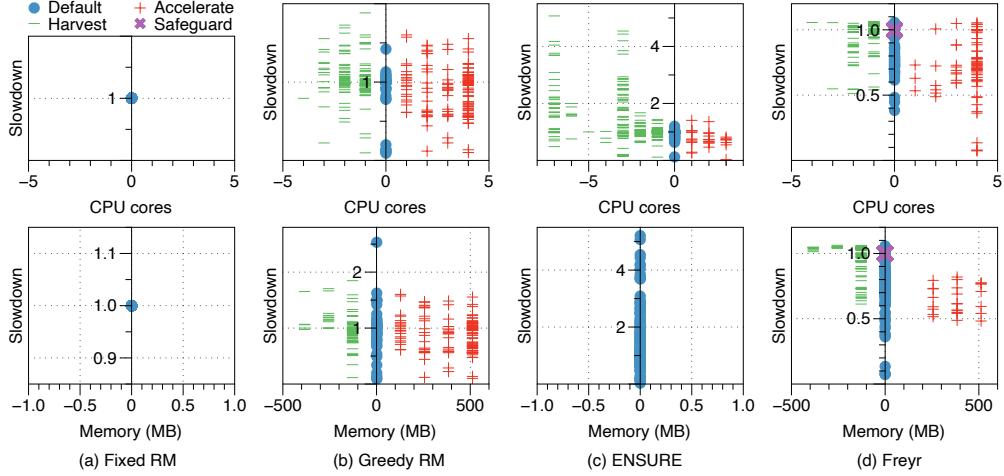}
    \vspace{-0.15in}
    \caption{Performance of individual invocations processed by Fixed RM, Greedy RM, ENSURE, and \sys in OpenWhisk evaluation. {Default ($\bullet$)}: invocations with user-requested allocation. {Accelerate ($+$)}: invocations accelerated by supplementary allocation. {Harvest ($-$)}: invocations with resource harvested. {Safeguard ($\times$)}: invocations protected by the safeguard.}
    \vspace{-0.15in}
    \label{fig:openwhisk-per-invocation}
\end{figure*}

We implement \sys with 6K lines of Scala code in Apache OpenWhisk \cite{openwhisk} and deploy it to a realistic OpenWhisk cluster. 
We train and evaluate \sys using realistic workloads from public serverless benchmarks and invocation traces sampled from Azure Functions traces~\cite{azure-traces} (implementation details in Supplementary Materials~\ref{sup:openwhisk}). 


\subsection{Methodology}
\label{sec:methodology}

\parab{\textbf{Baselines.}}
We compare \sys with three baseline RMs:
1) \textit{Fixed RM}: the default RM of most existing serverless platforms that allocates CPU cores in a fixed proportion to user-defined memory sizes. 
2) \textit{Greedy RM} detects a function's saturation points  based on its historical resource usage by gradually decreasing (increasing) the allocation for an over-provisioned (under-provisioned) function in a fine-tuned and fixed step. Our implementation sets the detect step size one core and 64 MBs for CPU and memory, respectively. Besides, Greedy RM allocates resources to functions in a first-come-first-serve manner.
%
3) \textit{ENSURE}~\cite{ensure} allocates memory resources as users request and adjusts the CPU cores for each function at runtime when detecting performance degradation.

\parab{\textbf{Evaluation metrics.}}
We use the \textit{slowdown} value defined in Section~\ref{subsec:formulation} to measure the performance of a function invocation. Function invocations with lower slowdowns have lower response latency.
For resource harvesting, \sys aims to maximize the amount of harvested resources while having minimal impact on the performance of victim functions. 
For resource re-assignment, \sys treats invocations with different lengths of response latency as the same by reducing slowdowns, which improves the overall performance of the workload.
We also report the details of SLO violation and 99th-percentile (P99) function response latency of the workload.

\parab{\textbf{Testbed:}}
We deploy and evaluate \sys on an OpenWhisk cluster with 13 physical servers. Two of the servers host the OpenWhisk components, such as the frontend, the controller, the messaging queue, and database services. One deploys the \sys agent. The remaining ten servers serve as the invokers for executing functions.
%
%
The server hosting \sys agent has 16 Intel Xeon Skylake CPU cores and 64 GB memory, and each of the other 12 servers has eight Intel Xeon Skylake CPU cores and 32 GB memory.
Each function can be configured with eight CPU cores and 1,024 MB of RAM at most.
Considering the serverless functions' short lifecycle, we monitor their CPU and memory usage per 0.1 second and keep the historical resource usage in the Redis (\ie, KV store in Figure~\ref{fig:openwhisk}). 

\parab{\textbf{Workloads:}}
We randomly sampled two function invocation sets for OpenWhisk evaluation. Table~\ref{table:workloads} depicts the two invocation sets (OW-train and OW-test) used in the OpenWhisk evaluation.
We use a scaled-down version of the invocation traces, \ie, we assume the invocation trace is based on seconds rather than minutes. 
This re-scaling increases the intensity of workloads while speeding up \sys OpenWhisk training by reducing the total workload duration.
We employ ten real-world functions from three serverless benchmark suites: SeBS \cite{copik2020sebs}, ServerlessBench \cite{serverlessbench}, and ENSURE-workloads \cite{ensure} (details of the ten functions in Supplementary Materials~\ref{sup:workload}).
For DH, EG, IP, KNN, ALU, MS and GD, each is initially configured with four CPU cores and 512 MB memory; for VP, IR and DV, each is initially configured with eight cores and 1,024 MB.
We set the initial resource configuration of each function according to the default settings from the suites. 

\begin{table}[b]
    \centering
    \vspace{-0.15in}
    \caption{Characterization of training and testing workload sets in the OpenWhisk evaluation. Metrics include: total number of unique traces, total number of invocations (Calls), average inter-arrival time (IAT), and requests per second.}
    \vspace{-0.1in}
    \begin{small}
        \begin{tabular}{cccccc}
            \toprule
            \textbf{Set} & \textbf{Traces} & \textbf{Calls} & \textbf{Avg IAT (s)} & \textbf{Reqs/sec} \\
            \otoprule
            OW-train     & 1,000             & 26,705         & 2.21                 & 0.44              \\
            OW-test      & 10             & 268            & 2.20                 & 0.45              \\
            \bottomrule
        \end{tabular}
    \end{small}
    \label{table:workloads}
\end{table}

\subsection{Results}

We summarize the slowdown and resource allocation of function invocations of the testing workload in Figure~\ref{fig:openwhisk-per-invocation}. 
In each subgraph, each point (\ie, $\bullet$, $+$, $-$, and $\times$) indicates a function invocation. 
The y-axis indicates the slowdown values of function invocations, and the x-axis shows the CPU and memory allocation of function invocations relative to their user configurations. The negative CPU and memory values indicates that RMs harvest corresponding resources from those invocations, and the positive means that those invocations are provided with additional resources.

\parab{\textbf{Overall performance.}}
\sys outperforms other baseline RMs with the best overall performance.
For processing the same testing workload, \sys achieves a lowest average slowdown of 0.82, whereas Fixed RM, Greedy RM, ENSURE are 1.0, 1.12, and 1.78, respectively.
Recall in Section~\ref{subsec:formulation}, a lower slowdown indicates a faster function response.
Compared to the default RM in OpenWhisk, \sys provides an average of 18\% faster function executions and 32.1\% lower P99 response latency for the testing workload. 
\sys harvests idle resources from 38.8\% of function invocations and accelerates 39.2\% of invocations.

\parab{\textbf{Harvesting and acceleration.}}
Figure~\ref{fig:openwhisk-per-invocation} shows the performance of 268 individual invocations processed by four RMs.
Fixed RM has no resource adjustment during its workload processing. 
Greedy harvests an average of 1.7 cores and 168 MB from victim invocations and accelerates under-provisioned functions with an average of 3 cores and 392 MB.
ENSURE's policy also harvests and accelerates invocations with CPU cores but makes no changes to memory resources.
ENSURE harvests an average of 3.4 cores from victims and accelerates under-provisioned functions with an average of 1.9 cores.
\sys harvests an average of 1.5 cores and 380 MB from victims and accelerates under-provisioned functions with an average of 3.6 cores and 164 MB.
\sys re-assign harvested resources to accelerate under-provisioned invocations, which speeds up for under-provisioned function invocations up to 92\%.

\parab{\textbf{SLO violation.}}
Figure~\ref{fig:openwhisk-per-invocation} shows that both Greedy RM and ENSURE severely violate function SLOs since there are some function invocations with slowdown values much larger than 1.
Fixed RM has no violation as it performs no harvesting or acceleration.
Greedy RM degrades the performance of some victim invocations over 60\%.
ENSURE violates SLOs of some victim invocations over 500\% when harvesting CPU cores.
Compared to Greedy RM and ENSURE, \sys rationally harvests idle resources from under-provisioned invocations, as the performance degradation of victim invocations is limited within 6\%.
When harvesting idle resources, \sys calls safeguard for 21.8\% of invocations to avoid potential performance degradation due to usage spike. 

\begin{figure}[t]
    \centering
    \begin{subfigure}[b]{0.22\textwidth}
        \centering
        \includegraphics[width=\textwidth]{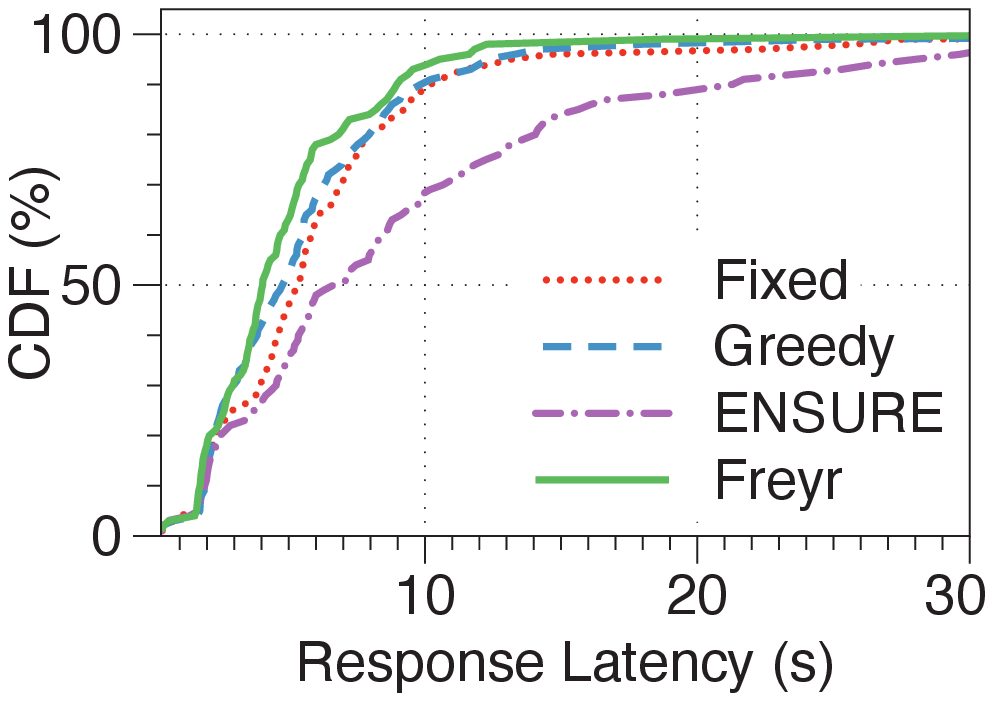}
        \caption{CDF of response latency}
        \label{subfig:openwhisk_cdf_latency}
    \end{subfigure}
    \begin{subfigure}[b]{0.22\textwidth}
        \centering
        \includegraphics[width=\textwidth]{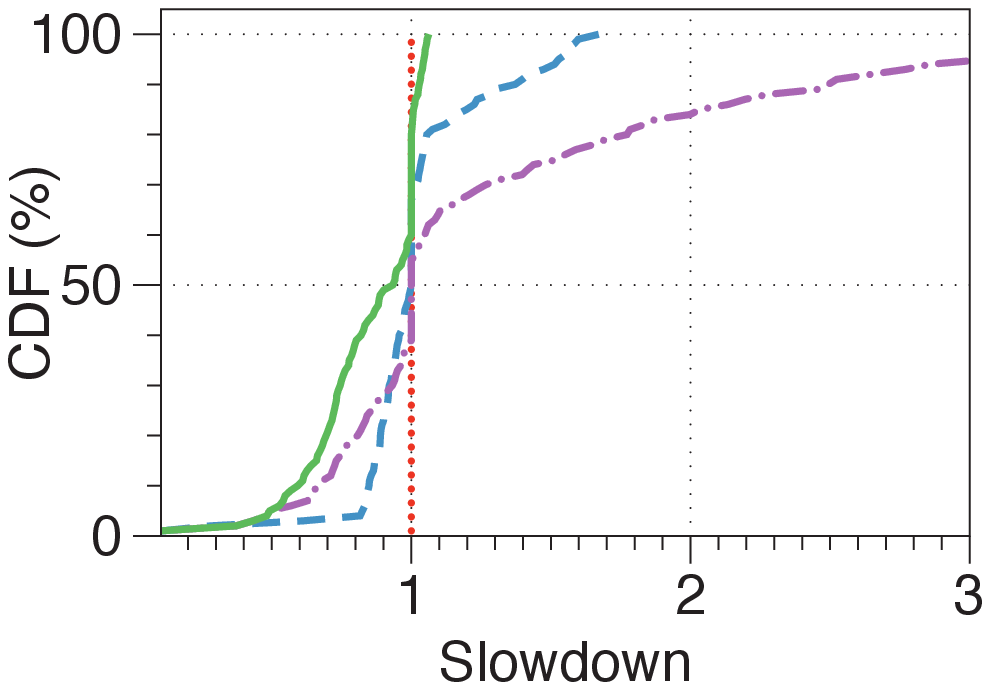}
        \caption{CDF of slowdown}
        \label{subfig:openwhisk_cdf_slowdown}
    \end{subfigure}
    \vspace{-0.1in}
    \caption{The CDF of function response latency (left) and slowdown (right) in OpenWhisk experiment respectively.}
    \label{fig:openwhisk_cdf}
    \vspace{-0.2in}
\end{figure}

\parab{\textbf{P99 latency.}}
Figure~\ref{subfig:openwhisk_cdf_latency} shows the CDF of function response latency of the testing workload. 
\sys has a P99 function response latency in less than 19 seconds, whereas Fixed RM, Greedy RM and ENSURE are 28, 25, and 38 seconds, respectively.
Figure~\ref{subfig:openwhisk_cdf_slowdown} shows the CDF of the slowdown of the testing workload. 
\sys maintains P99 slowdowns below 1.06 for all invocations, whereas Greedy RM and ENSURE are 1.58 and 4.5, respectively.
As Fixed RM adjusts no resources, the slowdown stays 1.0 for all percentile.


\section{Related Work}


\parab{Resource harvesting}. 
Research has been conducted on VM resource management in traditional clouds for years. SmartHarvest~\cite{wang2021smartharvest} proposes a VM resource harvesting algorithm using online learning. 
Unlike \sys, which uses harvested resources to accelerate function executions, SmartHarvest offers a new low-priority VM service using harvested resources. 
Directly replacing \sys with SmartHarvest is not feasible as SmartHarvest is not designed for serverless computing. 
Zhang~\etal~\cite{yanqizhang2021sosp} proposed to harvest VMs for serverless computing, while \sys harvests idle resources of serverless functions directly.

\parab{Resource provisioning}. 
Spock~\cite{gunasekaran2019spock} proposes a serverless-based VM scaling system to improve SLOs and reduce costs.
For resource management in serverless, \cite{cpu-cap} and \cite{ensure} both aim to automatically adjust CPU resource when detecting performance degradation during function executions, which help mitigate the issue of resource over-provisioning. 
Unlike \cite{cpu-cap} and \cite{ensure} that only focus on CPU, \sys manages CPU and memory resources independently. 
Kaffes~\etal\cite{core-granular} propose a centralized scheduler for serverless platforms that assigns each CPU core of worker servers to CPU cores of scheduler servers for fine-grained core-to-core management. 
\sys focuses on resource allocation rather than scheduling or scaling. 
Fifer~\cite{gunasekaran2020fifer} tackles the resource under-utilization in serverless computing by packing requests to fewer containers for function chains.
Instead of improving packing efficiency, \sys directly harvests idle resources from under-utilized functions.

\parab{Reinforcement learning}. 
\textsc{Siren} \cite{siren2019infocom} adopts DRL techniques to dynamically invoke functions for distributed machine learning with a serverless architecture. 
Our work \sys leverages DRL to improve the platform itself rather than serverless applications.
Decima~\cite{decima} leverages DRL to schedule DAG jobs for data processing clusters.
Metis~\cite{metis} proposes a scheduler to schedule long-running applications in large container clusters.
TVW-RL~\cite{mondal2021scheduling} proposes a DRL-based scheduler for time-varying workloads.
George~\cite{li2021george} uses DRL to place long-running containers in large computing clusters.
Differ from the above works, \sys learns resource management in serverless computing using DRL.




\section{Conclusion}
This paper proposed a new resource manager, \sys, which harvests idle resources from over-provisioned functions and accelerates under-provisioned functions with supplementary resources. Given realistic serverless workloads, \sys improved most function invocations while safely harvesting idle resources using reinforcement learning and a safeguard mechanism.
Experimental results on the OpenWhisk cluster demonstrate that \sys outperforms other baseline RMs.
\sys harvests idle resources from 38.8\% of function invocations and accelerates 39.2\% of invocations.
Compared to the default RM in OpenWhisk, \sys reduces the 99th-percentile function response latency by 32.1\% for the same testing workload.
\section{Acknowledgements}

This work is supported in part by the US National Science Foundation under grant number OIA-1946231, CRII-CNS-2104880, CRII-SaTC-1948374, and the Louisiana Board of Regents for the Louisiana Materials Design Alliance (LAMDA). Any opinion and findings expressed in the paper are those of the authors and do not necessarily reflect the view of the  funding agency.

\bibliographystyle{ACM-Reference-Format}
\bibliography{main}

\clearpage
\appendix

%


\section{The Training Algorithm}
\label{sup:training}

\sys uses a policy gradient algorithm for training.
Policy gradient methods are a class of RL algorithms that learn policies by performing gradient ascent directly on the parameters of neural networks using the rewards received during training.
When updating policies, large step sizes may collapse the performance, while small step sizes may decrease the sampling efficiency.
We use the Proximal Policy Optimization (PPO) algorithms \cite{ppo2} to ensure that \sys takes appropriate step sizes during policy updates. More specifically, given a policy $\pi_\theta$ parameterized by $\theta$,
the PPO algorithm updates policies at the $k$-th episode via
\begin{equation*}
    \theta_{k+1} = \mathop{\arg\max}\limits_{\theta} \mathop{\mathbb{E}}\limits_{s,a \sim \pi_{\theta_k}} \Big[\mathbb{L}(s,a,\theta_k,\theta)\Big],
    \label{eq:ppo-update}
\end{equation*}
where $\mathbb{L}$ is the \textit{surrogate advantage} \cite{trpo}, a measure of how policy $\pi_\theta$ performs relative to the old policy $\pi_{\theta_k}$ using data from the old policy. We use the PPO-clip version of a PPO algorithm, where $\mathbb{L}$ is given by
\begin{equation*}
    \mathbb{L}(s,a,\theta_k,\theta) = \min \Big(\frac{\pi_{\theta}(a|s)}{\pi_{\theta_k}(a|s)}A^{\pi_{\theta_k}}(s,a),\ g(\epsilon, A^{\pi_{\theta_k}}(s,a))\Big),
    \label{eq:ppo-adv}
\end{equation*}
and $g(\epsilon, A)$ is a clip operation defined as
\begin{equation*}
    g(\epsilon, A) =
    \begin{cases}
        (1+\epsilon)A, & \text{if $A \geq 0$}, \\
        (1-\epsilon)A, & \text{otherwise},
    \end{cases}
    \label{eq:ppo-g}
\end{equation*}
where $A$ is the advantage calculated as rewards $r$ subtracted by baseline values $b$; $\epsilon$ is a hyperparameter that restricts how far the new policy is allowed to deviate from the old.
Intuitively, the PPO algorithm sets a range for step sizes of policy updates, which prevents the new policy from deviating too much from the old (either positive or negative).

\begin{algorithm}[tb]
    \DontPrintSemicolon
    \SetNoFillComment
    \caption{\sys Training Algorithm.}
    \label{algo:freyr}
    Initial policy (actor network) parameters $\theta_0$ and value function (critic network) parameters $\phi_0$\;
    \For{episode k $\gets$ 0, 1, 2, \ldots}{
        Run policy $\pi_k = \pi(\theta_k)$ in the environment until $T$-th invocation completes\;
        Collect set of trajectories $\mathbb{D}_k = \{\tau_i\}$, where $\tau_i = (s_i, a_i), i \in [0, T]$\;
        Compute reward $\hat{r}_t$ via Equation~\ref{eq:rewards}\;
        Compute baseline value ${\Bar b}_t$ via Equation~\ref{eq:baseline-value}\;
        Compute advantage $\hat{\mathbb{A}}_t = \hat{r}_t - {\Bar b}_t$\;
        Update actor network by maximizing objective using stochastic gradient ascent:%
        \begin{align*}
                \theta_{k+1} = \arg\mathop{\max_{\theta}} \frac{1}{|\mathbb{D}_k|T} \mathop{\sum}_{\tau \in \mathbb{D}_k}\mathop{\sum}^T_{t=0} \mathbb{L}(s_t,a_t,\theta_k,\theta)%
        \end{align*}\;%
        Update critic network by regression on mean-squared error using stochastic gradient descent:%
        \begin{equation*}
            \phi_{k+1} = \arg\mathop{\min_{\phi}} \frac{1}{|\mathbb{D}_k|T} \mathop{\sum}_{\tau \in \mathbb{D}_k}\mathop{\sum}^T_{t=0} ({\Bar b}_t - \hat{r}_t)^2%
        \end{equation*}\;%
    }
\end{algorithm}

Algorithm~\ref{algo:freyr} presents the training process of \sys. 
For each episode, we record the whole set of trajectories including the states, actions, rewards, baseline values predicted by the critic network, and the logarithm probability of the actions for all invocations. 
After each training episode finishes, we use the collected trajectories to update the actor and critic networks.


\section{Implementation Details}
\label{sup:openwhisk}

Apache OpenWhisk is an open-source, distributed serverless platform that powers IBM Cloud Functions \cite{ibm-cloud-functions}. Figure~\ref{fig:openwhisk} illustrates the architecture of \sys based on OpenWhisk. OpenWhisk exposes an NGINX-based REST interface for users to interact with the platform. Users can create new functions, invoke functions, and query results of invocations via the frontend. The Frontend forwards function invocations to the Controller, which selects an Invoker (typically hosted using VMs) to execute invocations. The Load Balancer inside the Controller implements the scheduling logic by considering Invoker's health, available capacity, and infrastructure state. Once choosing an Invoker, the Controller sends the function invocation request to the selected Invoker via a Kafka-based distributed messaging component. The Invoker receives the request and executes the function using a Docker container. After finishing the function execution, the Invoker submits the result to a CouchDB-based Database and informs the Controller. Then the Controller returns the result of function executions to users synchronously or asynchronously. Here we focus on resource management for containers. 

We modify the following modules of OpenWhisk to implement our resource manager:

\textbf{Frontend:} Initially, OpenWhisk only allows users to define the memory limit of their functions and allocates CPU power proportionally based on memory. To decouple CPU and memory, we add a CPU limit and enable the Frontend to take CPU and memory inputs from users. Users are allowed to specify CPU cores and memory of their functions, and the Frontend forwards both CPU and memory limits to the Controller.

\textbf{Controller:} The Load Balancer makes scheduling decisions for the Controller. When selecting an Invoker, the Load Balancer considers available memory of Invokers. We modify the Load Balancer also to check available CPU cores of Invokers---the Load Balancer selects Invokers with enough available CPU cores and memory to execute function invocations.

\textbf{Invoker:} The Invoker uses a semaphore-based mechanism to control containers' access to available memory. We apply the same mechanism to control access to available CPU cores independently.

\textbf{Container:} By default, OpenWhisk uses \texttt{cpu-shares} parameter to regulate CPU power of containers. When plenty of CPU cycles are available, all containers with \texttt{cpu-shares} use as much CPU as they need. While \texttt{cpu-shares} improves CPU utilization of Invokers, it can lead to performance variation of function executions. We change the CPU parameter to \texttt{cpus} which restricts how many CPU cores a container can use. This is aligned with the CPU allocation policy of AWS Lambda \cite{awslambdalimits}. For each function invocation, we monitor the CPU cores and memory usage of its container using \texttt{cgroups}. We record the usage peak during function execution and keep it as history for \sys to query.

\textbf{DRL agent:} We implement the \sys's agent using two neural networks, each with two fully connected hidden layers.
The first hidden layer has 32 neurons, and the second layer has 16 neurons. 
Each neuron uses \texttt{Tanh} as its activation function. 
The agent is implemented in 2K lines of Python code using PyTorch \cite{pytorch}. 
\sys is lightweight because the policy network consists of only 1858 parameters (12 KB in total). 
Mapping a state to an action takes less than 10 ms.

\section{Workload Characterizations}
\label{sup:workload}

Table~\ref{table:functions} describes the type and dependency of 10 serverless applications from benchmark suites. 
DH downloads HTML template, populates the templates based on input, and uploads them to CouchDB. EG generates emails based on the input and returns them to the CouchDB. IP downloads images, resizes them, and uploads them to CouchDB. VP downloads videos, trims and tags them with a watermark, and uploads to CouchDB. IR downloads a batch of images, classifies them using ResNet-50, and uploads them to CouchDB. KNN downloads the dataset, performs the KNN algorithm on it, and uploads the result to CouchDB. GD performs three kinds of gradient descent based on input and uploads the result to CouchDB. ALU computes the arithmetic logic based on input and uploads the result to CouchDB. MS performs merge sorting based on input and uploads the result to CouchDB. DV downloads a DNA sequence file, visualizes the sequence, and uploads the result to CouchDB.
We profile the ten applications configured with eight CPU cores and 1,024 MB memory, which is the maximum allocation in our experimental environment.



\begin{table}[t]
    \centering
    \caption{Characterizations of serverless applications used in OpenWhisk evaluation. (DH: Dynamic HTML, EG: Email Generation, IP: Image Processing, VP: Video Processing, IR: Image Recognition, KNN: K Nearest Neighbors, GD: Gradient Descent, ALU: Arithmetic Logic Units, MS: Merge Sorting, and DV: DNA Visualization.)}
    \begin{small}
        \begin{tabular}{lccccc}
            \toprule
            \textbf{Function}           & \textbf{Type}    & \textbf{Dependency} \\
            \otoprule
            DH           & Web App          & Jinja2, CouchDB \\
            EG       & Web App          & CouchDB \\
            IP       & Multimedia       & Pillow, CouchDB \\
            VP       & Multimedia       & FFmpeg, CouchDB \\
            IR      & Machine Learning & Pillow, torch, CouchDB \\
            KNN   & Machine Learning & Scikit-learn, CouchDB  \\
            GD       & Machine Learning & NumPy, CouchDB \\
            ALU & Scientific       & CouchDB \\
            MS          & Scientific       & CouchDB  \\
            DV      & Scientific       & Squiggle, CouchDB  \\
            \bottomrule
        \end{tabular}
    \end{small}
    \vspace{-0.1in}
    \label{table:functions}
\end{table}

\section{Safeguard Sensitivity Analysis}
\label{sup:deep_dive}

\begin{figure}[t]
    \centering
    \includegraphics[width=\linewidth]{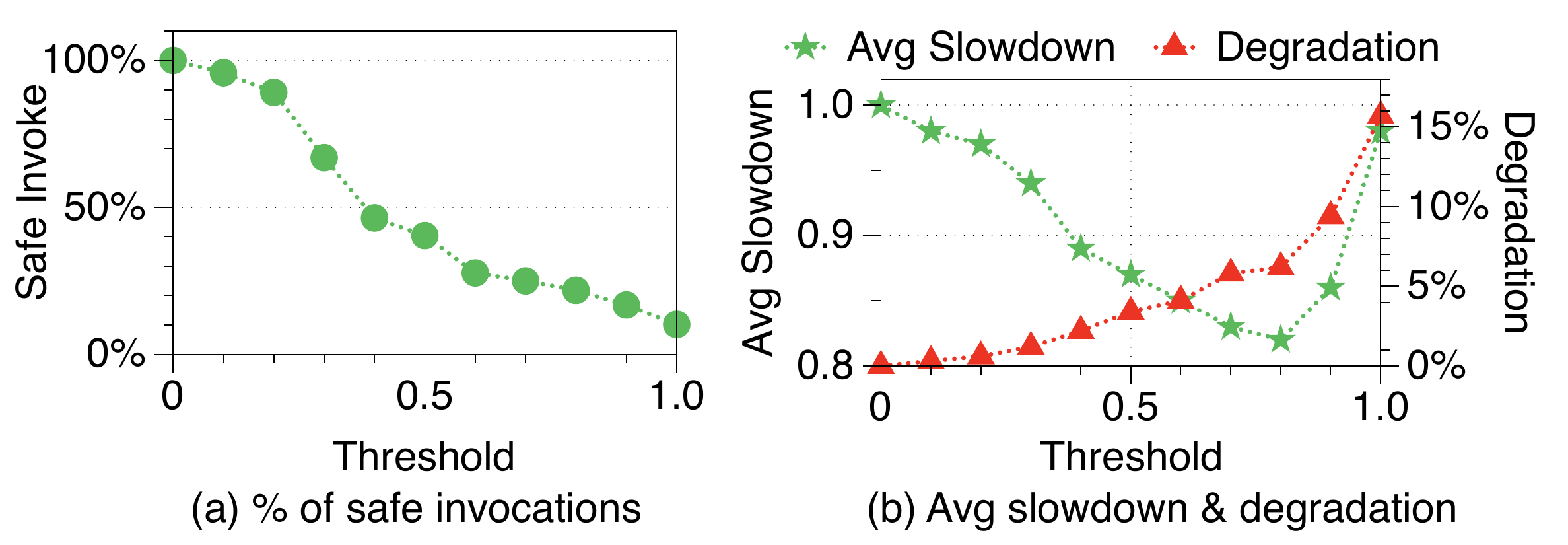}
    \Description{}
    \caption{Sensitivity analysis of safeguard thresholds.}
    \vspace{-0.1in}
    \label{fig:sensitivity}
\end{figure}


\parab{\textbf{Safeguard threshold.}}
We set the default threshold value in the safeguard algorithm to be 0.8, which allows \sys to trigger the safeguard just before detecting a full utilization.
The threshold is tunable---a high threshold may allow \sys to presumptuously harvest idle resource and deteriorate performance, while a low threshold may too conservatively restrict the harvesting and under-utilize resources.
We conduct a threshold analysis on our OpenWhisk testbed using the workload OW-test from Table~\ref{table:workloads} to evaluate the sensitivity of safeguard threshold in \sys.
We increase \sys's safeguard threshold from 0 to 1 with a step of 0.1 and run the same workload using \sys.
Figure~\ref{fig:sensitivity}(a) shows the percentage of safe invocations (invocations allocated with user-defined CPU/memory) under each threshold. 
Figure~\ref{fig:sensitivity}(b) shows the average slowdown and percentage of degraded invocations under each threshold.
When increasing the threshold, the rate of safe invocation drops down as \sys gradually harvests idle resources wildly. 
The percentage of degraded invocations gradually rises because \sys's harvesting policy becomes more and more unrestricted.
For average slowdown of the workload, \sys achieves better and better overall performance until its threshold reaching 0.8.
Due to severe performance degradation, \sys yields a worse performance for thresholds 0.9 and 1.0.

To deploy \sys in a production environment, service providers can tune the safeguard threshold based on their own criteria, \ie, tightening the threshold to conservatively harvest or loosing the threshold to actively harvest idle resources.


\parab{\textbf{Safeguard effectiveness.}}
To examine safeguard effectiveness in \sys, we also evaluate a variant of \sys with safeguard turned off.
We run the workload OW-test from Table~\ref{table:workloads} on our OpenWhisk testbed using safeguard-off \sys and obtain the average slowdown and performance degradation.
\sys without safeguard processes the testing workload with an average slowdown of 1.28 while degrading at most 15.7\% to function response latency, which is 36\% slower and has 9.5\% more degradation than the original version.
The result shows that \sys's safeguard effectively regulates the decision-making process, thus guaranteeing the performance of individual functions.

\section{Deploying \sys}
\label{sup:deploy-sys}

In industrial serverless computing environments, such as OpenWhisk, AWS Lambda, and Google Cloud Functions, integrating \sys lead to merits for both service providers and users. For service providers, \sys carefully harvests idle resources and reuses them to accelerate function invocations, which improves the overall serverless platform's resource utilization. For users who mistakenly configured insufficient resource allocation for their functions, \sys transparently brings potential performance protection (\ie, faster function executions) using harvested idle resources without violating other users' SLOs. 

\end{document}